\documentclass[amsmath,amssymb,lengthcheck,aps,prl] {revtex4}
\usepackage[colorlinks,allcolors=blue]{hyperref}
\usepackage[T1]{fontenc}
\usepackage[utf8]{inputenc}
\usepackage{graphics}
\usepackage{graphicx}
\begin{document}

\title{ Collective spin density excitations of fractional quantum Hall states in dilute ultracold Bose atoms}

\author{ Moumita Indra }
\author{ Dwipesh Majumder }
\affiliation{Department of Physics, Indian Institute of Engineering Science and Technology, Howrah, WB, India}

\begin{abstract}
We have studied collective spin density excitations of fractional quantum Hall effect (FQHE) in rotating Bose-Einstein condensation for the three filling fractions of first series of Jain's composite fermion sequences. We have considered, short-ranged contact interactions between the Bose atoms as well as long range Coulomb interactions to compare the nature of the spectras with FQHE of electrons. Using Monte Carlo method for finite but large number of particles, the lowest order collective modes of spin-reversed sectors is calculated here, by computing the energy differences of the respective excitons from the fully polarized ground states.

\end{abstract}
\maketitle

Strongly correlated many body systems can be studied with the help of weakly interacting quassi-particles taking part to the system. Fractional quantum Hall effect (FQHE) \cite{fqhe,QHE} is one such example, where composite fermion (CF) \cite{Jain_CF} is the natural quassiparticle. CFs are bound state of electron and even number of quantized vortices. In terms of CF's most of the strongly interacting FQH states of Jain series can be mapped into non-interacting integer quantum Hall states (IQHS). 

In recent years, rotating Bose-Einstein condensate (BEC) \cite{BEC} provide an interesting platform to study FQHE \cite{FQHE_BEC}. Two dimensional (2D) system of fast rotating BEC, confined in a harmonic trap in XY plane produces a ficticious magnetic field along Z-axis, i.e. perpendicular to the 2D plane as similar to the magnetic field in 2D electron system (2DES). It leads to the formation of Landau levels (LL). Then there is a possibility of the FQHE in rapidly rotating BEC of charged neutral dilute Bose gas \cite{CF_on_BEC}. 

Bosons may reside entirely in the lowest LL (LLL) at low density of atoms and high rotation. This problem is then well and truly ascertained only by the interactions between the particles since the kinetic energy is frozen. In a dilute and cold gas, only binary collisions at low energy are relevant and these collisions are characterized by a single parameter, the s-wave scattering length \cite{s_wave}. So, we have assumed a poor approximation for the interaction by considering short distances between the atoms, i.e. delta type interaction \cite{Delta_pot}.

Correlated states arising from interparticle interactions in dilute rotating confined atomic Bose gases can be described by non-interacting CFs of Bose particles---bound states of odd number of flux quanta (say $ p = 1,3,5, \cdots$) and Bose atoms \cite{CF_on_BEC, Hamiltonian}. This bound state of a boson and an odd number of flux quanta (which are akin to vortices in their topological properties) obeys fermionic statistics. The CFs of Bose atoms feel reduced amount of magnetic field, 
\begin{eqnarray}
B^* = B - p \rho \phi_0
\end{eqnarray}
where $B$ is the external applied magnetic field, $\rho$ is the number density of the Bose particle, $\phi_0$ is the magnetic flux quantum. 
In this diminished  magnetic field CF forms Landau level, called $\Lambda$-levels. 
The lowest LL filling fraction of boson and filling fraction of CF ($n$, an integer number of filled $\Lambda$-level) is given by
\begin{eqnarray}
\nu = \frac{n}{n p + 1}
\label{CF_seq}
\end{eqnarray}
With this prescription, we can see that, interacting bosons can behave as spinless fermions \cite{CP_statistics} and the filling fractions exactly follow the Jain sequence \cite{Jain_CF}.  \\
Regnault and Jolicoeur studied the ground state and low-lying excited states \cite{exact_12} of FQHE in the dilute limit of rotating BEC with small number of bosons by exact diagonalization.
It has been seen that the CF theory well agreed with the exact diagonalization result for the $\nu=1/2$ filling fraction \cite{FQHE_BEC}. We have seen that the other filling fractions $1/4$ and $1/6$ of the series $\nu=1/(p+1)$ also well agreed with the exact diagonalization (The result has not been shown here). So the CF wave function is a suitable wave function for the $\nu=1/(p+1)$ series. 
 Two-component BECs is formed from the trapping of magnetic hyperfine states of same atomic species and also from the trapping of mixtures of two different atomic species. Mixture of two condensates corresponding to two different spin states of Rubidium 87, $ \mid F=1, \; m=-1 >  $ and $ \mid F=1, \; m=0 >$ \cite{2BEC} can be created by the process of sympathetic cooling.
 Experiments were also done with two component condensate of Sodium (${Na}^{23}$) in the $ \mid F=1, \; m=1,0 > $  hyperfine states \cite{PRL'82}. They formed long lived metastable excited states composed of magnetic domains. Realization of a mixture of BEC of different atomic species for $K^{41}, Rb^{87}$ \cite{PRL'89, PRL'100} and $Rb^{85}, Rb^{87}$ \cite{PRL'101} is also done in experiment. 
  The two component system suggests that the more complex structure of FQHE including the spin polarized states may exist \cite{2Comp_BEC}. Here we have not considered the spin-polarization states rather we have considered spin-reversed excitations of a fully polarized FQHS of Bose atoms.

The collective excitations in the FQHS of electron system has been well studied for almost four decades. 
The natural collective excitations in FQHS of electon system is the CF excitons \cite{exciton}. The long range coulomb interaction gives the roton types of spectum for spin conserving excitations. The low energy excitation is the spin-wave (SW) excitation.
In the small wave vector limit ($k \rightarrow 0$), Larmor’s theorem stipulates that the SW energy is precisely equal to the bare Zeeman, for a conventional ferromagnet, such as the one at $\nu=1$ \cite{nu1} or at $\nu=1/3$ \cite{SW}, SW has positive dispersion with energy that increases monotonically with wave vector reaching a large wave vector asymptotic limit of particle and hole seperation energy with opposite spin.

Recently, D. Das {\textit{et. al.}} have studied the collective charge density excitation of FQH states of rotating BEC using CF theory for filling fraction $ \nu = 1/2 $ \cite{Physica_B}. But best of our knowledge, spin density excitation (SDE) of FQHE of Bose particles has not been studied till date. We are focussed here to study spin excitation spectra of this system. We have calculated energy spectra for $\nu=1/(p+1)$ filling fractions of Bose system using CF theory. In our calculation we have seen that the SW excitations also present in this system. We have considered different types of interactions in our study to investigate the SW nature of excitations.

\section{Interacting potential}

\begin{figure}
 \centering
   \includegraphics[width=10cm]{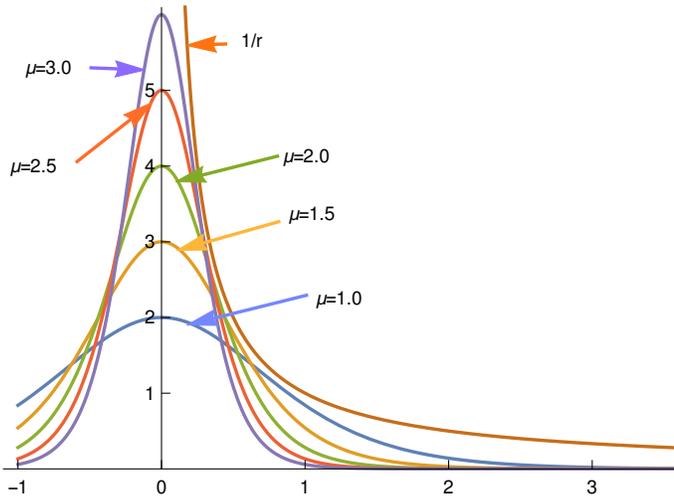}
   \caption{ P$\ddot{o}$schl-Teller interaction potential, $ V_{pt} = \frac{2 \mu}{cosh^2 \;( \mu r)}$ as function of separation distance ($r$) of two particles in unit of magnetic length ($l$) for different values of $\mu$ in unit of inverse of $l$ (data taken from Ref. \cite{Physica_B}).}
   \label{deltaPotl}
 \end{figure}
 
\subsection*{A. Delta function interaction }
Cooper, Wilkin and others applied the idea \cite{CF_on_BEC, Hamiltonian, BEC_FQHE} to map interacting bosons onto non-interacting spinless fermions and considered delta ($\delta$) function interaction between the Bose atoms.
For ultracold dilute bosons, the scattering between the atoms eventuates only in the s-wave. The effective interactions between the bosons at low energy limit, can be represented by a constant $U_ 0 = 4\pi  \hbar^2 a_s/m$ in momentum space, where $m$ is the mass of each particle and $a_s$ is the s-wave scattering length. 
 
This interaction gives the well known GP equation \cite{GP_equn} for BEC and superfluid system. In 2D 
the interaction potential is given by,
\begin{eqnarray}
V = g \sum_{i<j} \delta^{(2)}(\vec{r}_i - \vec{r}_j)
\end{eqnarray}
with, $g = \sqrt{32 \pi} \hbar \omega a_s l^2 / l_z $ \cite{exact_12}
Where, $l_z = \sqrt{\hbar/m \omega}$ is the characteristic length and $l$ is the magnetic length.
The interaction strength can be tuned by changing scattering length $a_s$ in presence of magnetic field. 

\subsubsection*{B. \textbf{P$\ddot{o}$schl-Teller interaction}}
It is very difficult to handle the $\delta$ function potential in quantum Monte Carlo calculation and it also requires huge computational resource. Using $\delta$ function potential it is not possible to calculate the energy spectra for large number of particles. To avoid this difficulty and to access the system size in the thermodynamic limit we have considered P$\ddot{o}$schl-Teller interaction potential ($ V_{PT} $) \cite{Poschl-Teller, Physica_B}.
\begin{eqnarray}
V_{PT} = U \sum_{i<j} \frac{2\mu}{cosh^2 \;( \mu r_{ij})}
\end{eqnarray}

where  $r_{ij}$ is the distance between two particles,  $\mu$ is the parameter of interaction in unit of inverse of the magnetic length $l$ i.e. 1/$\mu$ denotes the width of the interaction and $U$ is the interaction strength. Here we have considered a range of $\mu$, to investigate the $\mu$ dependence nature of excitation. As we increase the value of $\mu$ the nature of potential become $\delta$ type, as shown in FIG. \ref{deltaPotl}. The outmost line of this figure is the 1/r plot, which is shown to compare PT interaction with that of Coulomb. Very large value of $\mu$ will give zero energy as the average separation between particles will be large compared to the width of interaction.
So the beauty of this model potential is that we can change the range of interaction form short range $\delta$ function to long range Coulomb interaction by changing the parameter $\mu$.

 \begin{figure}
 \centering
   \includegraphics[width=0.45 \textwidth]{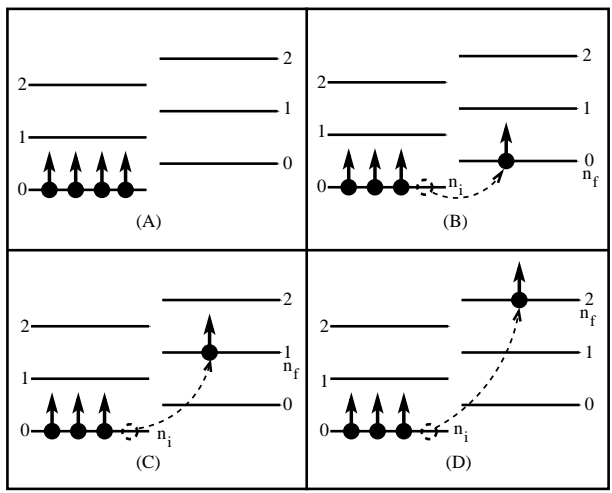}
   \caption{Schematic diagram of Spin-density excitons: In each block left panel represents up-spin $  (\uparrow )$ $\Lambda$-level, and right panel represents the down-spin $(\downarrow)$ $\Lambda$-level. Solid dot with arrow lines represents a CF, Bose atom (solid dot) with one magnetic flux attachment (arrow lines). The FQHE filling fraction of Bose particles $\nu=1/2, 1/4, 1/6$ maps into $n=1$ filled $\Lambda$-level (CF Landau level) with 1, 3, 5 flux attachment respectively. Spin reverse excitations in CF picture at $\nu=1/2$ are shown here, which follows for the other two filling fraction as well. Block (A) represents ground state and rest of the blocks (B), (C), (D) represent CF excitons (particle-hole pair), where dotted circle represents missing of particles from lowest spin $\uparrow$ $\Lambda$-level and that particle goes to $ n_f$-th spin $\downarrow$ $\Lambda$-level. Three lowest order excitations are considered here i.e. $\Longrightarrow $   (B)  0$\uparrow$ $\rightarrow$ 0$\downarrow$ ; (C) 0$\uparrow$ $\rightarrow$ 1$\downarrow$  ;  (D) 0$\uparrow$ $\rightarrow$ 2$\downarrow$. }
   \label{SDE}
 \end{figure}

 \subsubsection*{C. \textbf{Coulomb interaction}}
BEC of magnetically trapped alkali atoms was first acheived in 1995 by Davis and others \cite{Davis_1995}. After that, first FQHE of charged Bose system was studied by Cooper and Wilkin \cite{CF_on_BEC} with the model of non-interacting CF. Moreover, FQHE occurs in the 2DES under the strong repulsive Coulomb interaction, that is why we are also excited to see the collective excitation in FQHE in the long range 1/r interaction, although such kind of systems are rarely observed in nature.

\begin{eqnarray}
V_{Cou} = \sum_{i<j} \frac{C}{ r_{ij}}
\end{eqnarray}
where $C$ is the coefficient that contains all the information of charge and medium.

 \begin{figure}
 \centering
   \includegraphics[width=0.45 \textwidth]{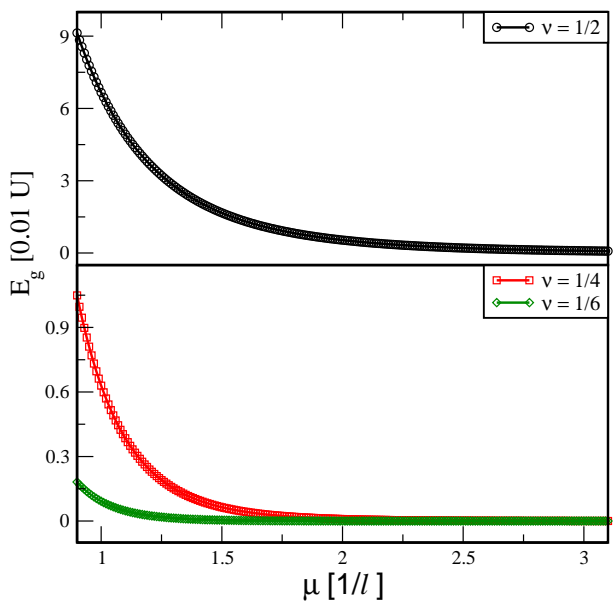}
   \caption{(Color online) Estimation of ground state energy per particles: We have calculated the ground state energy for several number of particles of the filling fractions $\nu = 1/2,1/4, 1/6 $. The average energy per particle is plotted as a function of the interaction parameter $\mu$.}
   \label{Eg}
 \end{figure}

 \begin{figure*}
 \centering
   {\includegraphics[width=0.68\textwidth]{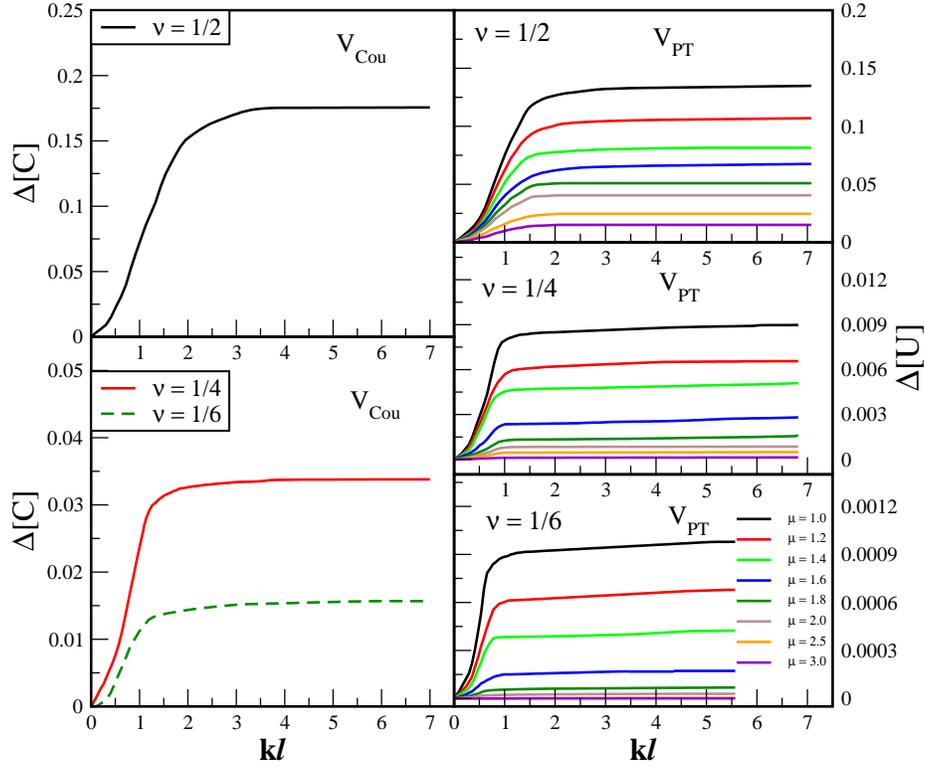}}
   \caption{(Color online) Spin density excitation: Energy spectra for different values of $\mu$  using PT- interaction potential for the three filling fractions $\nu = 1/2, 1/4, 1/6 $ are shown in the three right boxes. $\mu$ values are increasing downwards in each filling fractions. Energy spectra using Coulomb interaction potential are shown also in two left boxes. Wave vector ($k$) is related to the total angular momentum by $kl = L/R$, which is plotted along X-axis.}
   \label{Ex_SDE}
 \end{figure*}

\section*{Wave function \& calculation proccedures}

As the spherical geometry has no edge, it is beneficial to study the bulk properties of FQHE with finite number of electrons. In our numerical calculations, we thus formulate composite fermion wave function in spherical geometry \cite{Haldane, Jain_book}. It is thought that, $N$ number of correlated electrons are moving on the surface of a sphere, subjected to a radial magnetic field. The magnetic field is assumed to emerge from a ‘magnetic monopole’ of strength Q at the centre of sphere, which produces a total magnetic flux of $2Q\phi_0$ through the surface of the sphere of radius $R = \sqrt{Q} l$. This maps into a system of composite fermions at an effective flux $ 2q = 2Q - p(N - 1)$, with $Q$ chosen so that the state at $q$ is an integral quantum Hall state at integer filling $n $ so that, we will have $\nu = n/(np+1) $ filling fraction. The monopole charge $Q$ must be an integer or half integer according to Dirac’s quantization condition \cite{Dirac}. In spherical geometry, the angular momentum number is a good quantum number and its value of an electron in the $n$-th LL is $n+Q-1$ \cite{monopole_harmonics}. The single particle basis state is known as monopole harmonics which is given by \cite{Jain_kamilla},

\begin{eqnarray}
Y_{Q,n,m}(\Omega)&=& N_{Qnm} e^{iQ\phi} u^{Q+m} v^{Q-m} \\ \nonumber
&&\sum_{s=0}^{n}(-1)^s {{n \choose s}} {{ 2Q+n \choose Q+n-m-s}}  \\ \nonumber
 && (v^*v)^{n-s}(u^*u)^s\;\;
\label{mh}
\end{eqnarray}

where $u=cos(\theta/2) \; exp(-i\phi /2)$ and $v=sin(\theta/2) \; exp(i\phi /2)$ are the spinor variables with $0\le \theta \le \pi$ and $0\le \phi \le  2\pi$.

The ground state wave function of the $N$ electron system of FQHE at filling fraction $\nu$, which maps with $n$ filled $\Lambda$-levels of CFs is \cite{FQHE_BEC, wave_funcn},
\begin{equation}
  \Psi^0 = J^{-1} P_{LLL} J^2 \; \Phi_1(\Omega_1, \Omega_2, \cdots \Omega_N)
\end{equation}
where $\Omega_i$ are the position of electron on the surface of the sphere, $\Phi_1$ is the Slater determinant of completely filled lowest $\Lambda$-level of CFs, $P_{LLL}$ is the projection operator onto the LLL \cite{PLL} and the Jastrow factor is given by
\begin{eqnarray}
  J = \prod _{i<j}^N (u_i v_j - u_j v_i)^p  \nonumber
\end{eqnarray}
Here $\Phi_1$ and $J$ both are odd under the exchange of particles, so the total wave function of the system remains symmetric after the fermionic transformation. \\

 In case of SDE study, the excited state wave function for spin-reversed $N$-particle system corresponding to the transition of a CF from the filled  $\Lambda$-level to an other spin $n_f\;$-th $\Lambda$-level is obtained by \cite{DM_SSM},

{\small{
\begin{eqnarray}
  \Psi (L) &=& J^{-1} P_{LLL} J^2 \sum_{m_h} |m_h>_{\small{N-1}}\; Y_{q,n_f,m_p}  \nonumber\\
  && <q, m_h; n_f+q,m_p|L,M>
\end{eqnarray}}}

  where, $|m_h>_{\small{N-1}}$ is the Slater determinants of $N-1$ number of particles of the filled lowest $\Lambda$-level with a hole at $m_h$ ($m_h$ is the Z-component of angular momentum) 
and $Y_{q,n_f,m_p} $ is one single particle state, which is nothing but the quassi-particle in another spin state of the $n_f\;$-th $\Lambda$-level with Z-component of angular momentum $m_p$. \; $<q, m_h;n_f+q,m_p|L,M>$ are the Clebsch-Gordan coefficients, $L$ is the total angular momentum. We have considered total Z-component of angular momentum zero ($M=0$), without any loss of generality to avoid the numerical complicacy.
 
 The transition 0$\uparrow$ $\rightarrow$ 0$\downarrow$ is the conventional SW excitation, whereas 0$\uparrow$ $\rightarrow$ 1$\downarrow$ and 0$\uparrow$ $\rightarrow$ 2$\downarrow$ are the spin-flip excitations, i.e. the spin-reversed excitations can be realized as one CF jumps from 0$\uparrow$ to any one of the spin down  $\Lambda$-levels.

 In FIG. \ref{SDE} possible collective SDEs are shown for $\nu = 1/2 $ filling fraction. Likewise, for the other filling fractions such kind of excitations is considered. We have presented the results of possible excitons as we have checked that this energy is identical with the calculation considering lowest excitons. The excitons are not orthogonal, we have used Gram-Schmidt Orthonormalization procedure \cite{GS_ortho} to orthogonalize low energy exciton states with a fixed angular momentum. \\
 
 The ground state energy is obtained using the variational principle as, 
 \begin{equation}
  E_g = \frac{<\Psi^0| H |\Psi^0 >}{<\Psi^0 | \Psi^0 >}
\end{equation}
 
The excited state energy with respect to the  ground state  $E_g$  is given by 
\begin{equation}
  \Delta(L) = \frac{<\Psi(L)| H |\Psi(L)>}{<\Psi(L) |\Psi(L)>} - \frac{<\Psi^0| H |\Psi^0 >}{<\Psi^0 | \Psi^0 >  }
\end{equation}
$H$ is Hamiltonian of the system. As the kinetic energy become quantized and we assume that the particles are confined in the LLL, the Hamiltonian of the system will be $H=V$. The multidimensional integration has been carried out using quantum Monte Carlo method.

\section*{Results \& discussion} 

Previous studies by Chang \textit {et al.} \cite{FQHE_BEC} show that the CF description of Bose atoms worsens with increasing $ n $ along the CF sequence [equation (\ref{CF_seq})] and the mapping is also quantitatively very accurate for the ground state and excited state at $ \nu = 1/2 $. We  have also seen that the CF results well agrees with the exact diagonalization for $\nu = 1/4, 1/6$ filling fractions. So the CF wave function is very good wave function for $\nu=1/(p+1)$ states. 
That is why we have studied here the spin-reversed collective excitations of the $\nu=1/(p+1)$ filling fraction of bosonic system using CF theory.

In this study we have considered Coulomb interaction as well as PT interaction.  We have a freedom to control the range of interaction in the PT model interaction, so that we can study the nature of excitations for diffrent range of interaction.   In FIG. \ref{Eg} variation of ground state energies ($E_g$) for different filling fraction as a function of $\mu$ have been shown for PT interaction potential. The energy reduces with the increase of the interaction parameter $\mu$, as the range of interaction reduces with $\mu$.  

We have calculated the energy spectra for different number of particles from 80 to 160 for each filling fraction. We have shown the average energy spectra in the FIG. \ref{Ex_SDE}. 
The figure clearly shows the spin-wave excitations similar to the fermionic FQHE. The interesting fact is that the nature of spectra does not depend on the range of interaction. From FIG. \ref{Ex_SDE} it is obvious that the energy should be reduced at shorter range of interaction. 
So we conclude that the spin-reversed excitations at filling fraction $\nu=1/(p+1)$ supports the SW excitations for all range of interactions.


\section*{Acknowledgement}

It is our pleasure to thank Sutirtha Mukherjee for sharing his exact diagonalization result with us. Moumita thanks DST INSPIRE (Ref: IF160850), India for the financial support.

\end{document}